\documentclass[preprint,aps,superscriptaddress]{revtex4}%12pt]{article}
\usepackage{graphicx}
\usepackage{amsmath}
\usepackage[usenames]{color}
\usepackage{amssymb}
\newcommand{\be}{\begin{equation}}
\newcommand{\ee}{\end{equation}}
\newcommand{\bea}{\begin{eqnarray}}
\newcommand{\eea}{\end{eqnarray}}
\newcommand{\pr}{\partial}
\newcommand{\nno}{\nonumber}
\newcommand{\bse}{\begin{subequations}}
\newcommand{\ese}{\end{subequations}}

\begin{document}
\title{Cosmological Behavior of a Parity and Charge-Parity Violating Varying Alpha Theory}
\author{Debaprasad Maity
\footnote{debu.imsc@gmail.com}}
\affiliation{Department of Physics and Center for
Theoretical Sciences, National Taiwan
University, Taipei 10617, Taiwan}
\affiliation{Leung Center for Cosmology and Particle Astrophysics\\
National Taiwan University, Taipei 106, Taiwan}
\author{Pisin Chen\footnote{chen@slac.stanford.edu}}
\affiliation{Department of Physics and Center for 
Theoretical Sciences, National Taiwan
University, Taipei 10617, Taiwan}
\affiliation{Leung Center for Cosmology and Particle Astrophysics\\
National Taiwan University, Taipei 106, Taiwan}
\affiliation{Kavli Institute for Particle Astrophysics and Cosmology\\
SLAC National Accelerator Laboratory, Menlo Park, CA 94025, U.S.A.}
%\maketit

\begin{abstract}
In this paper we construct a phenomenological model in which the time variation of the fine structure constant, $\alpha$, is induced by a parity and charge-parity (PCP) violating interaction. Such a PCP violation in the photon sector has a distinct physical origin from that in the conventional models of this kind. We calculate the cosmological birefringence so induced in our model and show that it in turn produces a new non-vanishing multipole moment correlation between the temperature and the polarization anisotropies in the CMB spectrum. We have also
calculated the amount of optical rotation due to a strong background magnetic field and the effect of our new PCP violating term on the variation of $\alpha$ during the cosmic evolution. We found that only in the radiation dominated era can the contribution of the new PCP violating term to the variation of $\alpha$ be non-vanishing. 
\end{abstract}

\maketitle

\section{Introduction}\label{intro}
Both inflation and late-time cosmic acceleration have puzzled
 physicists for a long time. It has become clear that 
the final solutions to these may require new physics beyond general relativity and the
standard model of particle physics in order to explain those 
observation. {\em A priori}, however, we do not have any clear
idea about how to proceed unless we can identify some new
guiding principles. Although several new principles, such 
as the holographic principle, have been introduced to 
explain cosmological phenomena, these are nonetheless 
still at the preliminary stage. An alternative would be 
the more conservative path of drawing an analogy from known physics. 

As is well-known, in parity (P) and charge-parity (CP) 
symmetries are violated in the electroweak sector of the 
standard model particle physics. Considering this as a 
guiding principle, construction of a P and CP violating extension has been
considered in the new physics models that produce inflation 
as well as late-time acceleration. 
For the last several years, many different parity 
violating models have been put forward \cite{carroll,marc,contaldi,soda,debu}.
The very basic idea of all those models is to add an explicit  
parity violating term in the Lagrangian. Because of its
nature, this parity violating term leads to cosmic 
birefringence \cite{carroll,marc} and left-right asymmetry
in the gravitational wave dynamics \cite{contaldi,soda}. 
String inspired models with non-standard parity-violating interactions 
have also been discussed \cite{debu}. 
Various observable effects of these new parity violating models have been extensively investigated in order to put constrains on the corresponding 
parameters. 

In this paper we construct a parity and 
charge-parity (PCP) violating model
in the framework of ``varying-alpha theory". Some aspects of our model are similar to that proposed by Carroll \cite{carroll}.
But as we will see, our model has the advantage over 
Carroll's in that the origin of the parity 
violation may be more physically motivated. 

Cosmological variation of fundamental 
constants in nature has gained considerable interests in the 
recent past because of
two fundamental reasons. First, 
triggered by the string theory there has been a resurgence
of motivation to reconsider the variation of fundamental constants
in cosmology as well as particle physics model building.  
As is well known, string theory gives us a consistent framework, where the
effective four dimensional fundamental constants depend on the compactifications of the
extra dimensions. In principle, therefore, all the
so-called fundamental constants in our four dimensional world 
could actually be spacetime varying functions. 
The dynamics of such varying  `constants' actually depends on 
the specific compactification that we make. Second, 
increasingly high precession cosmological as well as laboratory 
experiments give us 
hope that the signature of new physics, including those that 
give rise to the variation of fundamental constants, 
may emerge in the near future.

In spite of the long history of the speculation  
of the variation of fine structure constant \cite{gammow}, the first
consistent, gauge invariant and Lorentz invariant, framework of $\alpha$
variability was proposed by Bekenstein \cite{bek2}.
Subsequently this subject has attracted much 
attentions and it was extensively studied
in \cite{bsm,bsm1,barrow}, mainly due to the first observational
evidence from the quasar absorption spectra that the fine
structure `constant' might change with cosmological time
\cite{murphy,webb,webb2}. This observation suggests that the value of
$\alpha $ may be lower in the past in cosmological time scale,
with $\Delta \alpha/\alpha =-0.72\pm 0.18\times 10^{-5}$ for
redshift $z\approx 0.5-3.5.$

We organize this paper as follows: in Sec.\ref{sec1} we 
construct the PCP violating model in the photon sector after
briefly reviewing the basic concept of ``varying-alpha theory". 
Then we discuss the theoretical implication and prediction of our
model in different cosmological phenomena. In Sec.\ref{sec2} we
study the cosmic birefringence phenomena and calculate
the rotation angle of the polarization of the 
electromagnetic wave in a leading-order approximation. We then
discuss its effect on the parity violating 
correlation function in the CMB polarization spectrum. 
In Sec.\ref{sec3} we discuss
the effect of the background magnetic field on the rotation
of the plane of polarization. As we know,
there exists magnetic fields at cosmic scales that may affect the CMB 
polarization due to some scalar field coupling.
There exists several laboratory-based experiments that aim at measuring the change of 
polarization of a electromagnetic wave induced by 
such a non-trivial (pseudo) scalar-photon coupling in a 
background magnetic field. 
Motivated by all these, we calculate the amount of optical rotation induced in a background cosmic magnetic field,  
which has a direct contact with experiments.  
In the subsequent Sec.\ref{sec4} we first briefly review
the varying alpha cosmology and then calculate
the alpha variation induced by the PCP violating term. 
In general, it is very difficult to solve the type of 
equation of motion that appeared in our model.
This was done in our calculation by using the matched approximation
adopted from \cite{bsm1}. Concluding remarks and 
future prospects are provided in Sec.\ref{con}.

\section{Parity violating varying-alpha theory}\label{sec1}
In this section we will start with a general discussion 
on the varying fine structure constant theory from the standard literatures 
\cite{bek2,bsm,bsm1} .
In the framework of the varying alpha theory, 
the simplest way to induce the variation of $\alpha$ is by requiring that
the electric charge varies as $e=e_{0} e^{\phi(x)}$, where 
$e_0$  denotes the coupling constant of a particle and $\phi(x)$ 
is a dimensionless scalar field. The fine-structure constant is therefore $\alpha = e_0^2 e^{2\phi(x)}$. 
There is an arbitrariness involved in the
definition of $\phi(x)$ due to the shift invariance,
i.e. $\phi\rightarrow \phi + c$.  
An important point to mention here is that the well known charge conservation 
is violated. But in order to be consistent with the quantum field theory,
new modified electromagnetic theory should be gauge invariant. 
Since $e$ is the electromagnetic coupling, the $\phi(x) $ field 
couples to the gauge field as $e^\phi(x) A_\mu $ in the 
Lagrangian and in the gauge transformation which leaves the 
action invariant is 
\bea
e^\phi A_\mu \rightarrow e^\phi A_\mu +\chi_{,\mu }.
\eea
So, from the above considerations, the unique gauge-invariant 
and shift symmetric Lagrangian for the modified electromagnetic field 
can be written as 
\bea
S_{em}= - \frac 1 4 \int d^4 x \sqrt{-g} e^{-2\phi}  F_{\mu\nu}F^{\mu\nu},
\eea 
where the new electromagnetic field strength tensor is defined as
\bea
F_{\mu \nu }=(e^\phi A_\nu )_{,\mu }-(e^\phi
A_\mu )_{,\nu }.\eea
In the above action and for the rest of this paper we set $e_0=1$ for convenience.
As one can see, the above action reduces to the usual form when $\phi$ is constant. 
The dynamics of the $\phi(x) $
field is controlled by the kinetic term 
\bea
{\cal L}_\phi =- \frac {\omega}2 \int d^4 x \sqrt{-g}  \phi_{,\mu }\phi^{,\mu },
\eea
which is clearly invariant under the shift symmetry of $\phi$.
Here the coupling constant $\omega$ can be written as 
$\hbar c/l^2$, where $l$ is the
characteristic length scale of the theory above which the Coulomb force law
is valid for a point charge. From the present experimental 
constraints the energy scale, $\hbar c/l,$ has to be above a few tens of MeV
to avoid conflict with experiments.

One of the natural assumptions in constructing the above Lagrangian is 
time-reversal invariance. But we will relax this 
assumption and try to analyze its
implications. An obvious term that is consistent with the varying alpha
framework yet violates PCP is ${\tilde F}_{\mu\nu} F^{\mu\nu}$, 
where ${\tilde F}^{\mu\nu} = \epsilon^{\mu\nu\sigma\rho} F_{\sigma\rho}$
is the Hodge dual of the Electromagnetic field tensor. In the conventional
electromagnetism this does not contribute to the classical equation of motion.
But in the present framework this is no longer true because of its
coupling with the scalar field $\phi(x)$. As we have explained 
in the introduction, at the present level of experimental 
accuracy PCP violation in the electromagnetic sector may not be 
ruled out, and if the PCP in this EM sector is indeed violated, 
then there should be some interesting 
consequences. Motivated by this, we write down a parity violating
Lagrangian
\be
{\cal L} ~=~  M_p^2 R - \frac {\omega} 2 \pr_{\mu}\phi 
\pr^{\mu} \phi ~ - \frac 1 4 e^{-2\phi } 
F_{\mu\nu} F^{\mu\nu} 
+ \frac {\beta}{4}e^{-2\phi } F_{\mu\nu} {\tilde F}^{\mu\nu}~+~ 
{\mathcal L}_m,    \label{action}
\ee
where $R$ is the curvature scalar and $\beta$ is
a free coupling parameter in our model.
As we can see, the scalar field $\phi$ plays a
similar role as that of the dilaton in the 
low-energy limit of string and M-theories, the important 
difference being that it induces a PCP violating electromagnetic 
interaction. For our purpose, we assume $\beta$ as a free but small parameter.
Here we want to emphasize that the model can be thought of as
a unified framework for dealing with different cosmological phenomena. 
At the present level of experimental accuracy,  
investigations of parity or charge-parity violation, 
beyond-standard model may shed some new light 
on the fundamental laws of physics. 
With the interest of phenomenological impacts on the present cosmological
observations, subsequently we will discuss some consequences
of our model.   

Before this let us write down the full set of equations of motion 
\begin{equation}
G_{\mu \nu }= \frac 1 {M_p^2} \left(T_{\mu \nu }^{mat} 
~+~T_{\mu \nu }^{\Phi} ~+~e^{-2 \phi} T_{\mu \nu}^{em} \right) ,
\end{equation}
where the energy-momentum tensors are
\bea
&(a)& T_{\mu\nu}^{mat} = \frac 1 2 g_{\mu\nu} {\cal L}_m - \frac {\pr {\cal L}_m}{\pr g_{\mu\nu}},\\
&(b)& T_{\mu\nu}^{em} = \frac {1} 2 e^{-2 \phi} 
\left\{ F_{\mu\alpha} F_{\nu}^{\alpha} - 
\frac 1 4 g_{\mu\nu} F_{\mu\nu}F^{\mu\nu}\right\},\\
& (c)& T_{\mu\nu}^{\Phi} = \frac {\omega} {2} 
\left\{ \pr_{\mu}\phi \pr^{\nu}\phi - \frac 1 2 g_{\mu\nu} 
\pr_{\alpha}\phi \pr^{\alpha}\phi \right\}.
\eea
The electromagnetic field equation then becomes
\be
\frac 1 {\sqrt{-g}} \pr_{\mu}(\sqrt{-g}  F^{\mu\nu}) + \pr_{\mu}\phi
(- F^{\mu\nu} + \beta \tilde{F}^{\mu\nu}) = 0.
\ee
Varying it with respect to $\phi$, we get
\begin{equation}
\frac 1 {\sqrt{-g}} \pr_{\mu}(\sqrt{-g} \pr^{\mu} \phi) = \frac {e^{-2 \phi}}
{2 \omega}\left[- F_{\mu \nu }F^{\mu \nu} ~+~ 
\beta F_{\mu \nu } \tilde{F}^{\mu \nu} \right].  
\end{equation}
In the subsequent sections we will study some cosmic phenomena
which may be relevant to the future precision cosmological measurements.

\section{Cosmological Birefringence} \label{sec2}
Cosmological birefringence (CB) is a wavelength-independent
rotation of photon polarization vector after traversing
a long cosmic distance. It has long been the subject of interest  
in the context of cosmic microwave background (CMB) phenomena
\cite{carroll,marc,maroto,mas} where its polarization properties
crucially depend on CB. 
The origin of this effect may come from either cosmic inhomogeneities or 
some non-trivial coupling of photon with other fields. 
In this section, we will study this effect and show that the
main contribution to CB comes from our PCP violating
term in the Lagrangian in Eq.\ref{action}. In order to calculate
this effect, we assume the background spacetime as the spatially flat 
FRW expanding background. On that
background we will compute the cosmic optical rotation which is the 
measure of CB.
For this it useful to take the background FRW metric in the conformal time that is 
\bea
ds^2 = a(\eta)^2 (- d\eta^2 + dx^2 + dy^2 + dz^2),
\eea
where $\eta$ is the conformal time and $a(\eta)$ is 
the conformal scale factor. Since the electromagnetic theory 
is conformal invariance in four dimension, 
the Maxwell equations turn out to be of the standard type 
with the modifications coming from the non-trivial 
scalar field $\phi$ coupling .
\bea
\nabla \cdot {\bf E} &=&2 \nabla { \phi} \cdot {\bf E} - 
4 \beta \nabla { \phi} \cdot {\bf B} , \nonumber \\
\partial_{\eta}({\bf E}) - \nabla \times {\bf B} &=&
2 ({\dot { \phi}} {\bf E}-\nabla { {\phi}}\times {\bf B})
-4 \beta( {\dot{{\phi}}}
{\bf B} + \nabla {{\phi}} \times {\bf E}),\nonumber \\
\bf \nabla \cdot \bf B &=& 0,\nonumber \\
\partial_{\eta}\bf B + \bf \nabla \times \bf E &=& 0.
\eea
The wave equation for the ${\bf B}$ then becomes,
\bea
\ddot{\bf B} - \nabla^2 {\bf B} = 
2\dot{{\phi}} (
 \dot{\bf B}+ 2 \beta \nabla \times {\bf B}) .
\eea
Now, what is important point to keep in mind here is that
the definition of physical electromagnetic field strength we use are  
$F_{i0}= {\bf E}_i$ and 
$ F_{ij}= \epsilon_{ijk} {\bf B}_k$. Where, $i=1,2,3$ and $\epsilon$
is the three spatial dimensional Levi-Civita tensor density. 

We assume general wave solutions of the form
$
{\bf B}= {\bf B}_0(\eta) e^{-i {\bf k}\cdot {\bf z}},
$ and 
take the z direction as the propagation direction of the
electromagnetic waves, i.e.,${\bf k} = k {\hat {\bf e}}_z$,
The equations for the polarization states, viz.,
$b_{\pm}(\eta) = {\bf B}_{0x}(\eta)\pm i {\bf B}_{0y}(\eta) $ turns out to be
\bea
\ddot{b}_{\pm} -  2\dot{{\phi}} \dot{b}_{\pm} + 
\left( {\bf k}^2 \mp 
4 {\bf k} \beta \dot{{\phi}}\right){b}_{\pm}=0,
\eea
while the equation of motion for the scalar field is
\bea
\ddot{\phi} - 2 \frac {\dot{a}} {a} \dot{\phi} = \frac {1}
{\omega a^2} [
-( {\bf E}^2 - {\bf B}^2) + 4 \beta {\bf B}\cdot{\bf E}].
\eea
It is in general difficult to solve the above non-linear coupled equations exactly.
We therefore look for an approximate solution to the leading order in the 
large $\omega$ 
limit. In this limit, the solution for the scalar field would be 
\bea
\phi = B \int \frac {d\eta}{a(\eta)^2} + C + {\cal{O}}(\omega) ~~;~~
\dot {\phi}= \frac B
{a(\eta)^2}
\eea
where $B$ and $C$ are the integration constants. We also assume
the coupling constant $\beta$ and the value of the scalar field to be vary small
based on the various observational constraints.
From the above expressions, we see that the energy density of the scalar field
is proportional to $B$. We therefore know that this constant must be very small
in order for it not to backreact to the background cosmological evolution.

Since the change of $b_{\pm}$ is expected to be small, we estimate the optical activity using  the WKB method \cite{dmssssg}. In the long wavelength limit and for
small coupling constant $\beta$, we assume the solution
of the above equation for $b_{\pm}$ to be
\bea
b_{\pm}= e^{i k S_{\pm}(\eta)}~~~~;~~~~ S_{\pm}(\eta)= S_{\pm}^0 + \frac
1 k S_{\pm}^{1} + \dots 
\eea
Therefore the solution based on the above ansatz is
\bea
S_{\pm}^0 = \eta~~~;~~~S_{\pm}^1 =- \frac 1 2 (2 i  \pm 4 \beta) 
 \int  {\dot{\phi}} d\eta .
\eea
It is clear from the above solution that the expression for 
the optical rotation of the plane of polarization is 
\bea
\Delta = 4 \beta  \int_{\eta_i}^{\eta_f}  {\dot{\phi}} d\eta
 = 4 \beta |\phi(\eta_f)
- \phi(\eta_i)|,
\eea
where $\eta_i$ and $\eta_f$ are the initial and final conformal time for the 
electromagnetic field to be detected.
As expected, the leading contribution to the
cosmic optical rotation comes from the PCP violating term.
In order to connect with observations, we rewrite the above expression 
for the optical rotation to the leading order in $\omega$ as
\bea 
\Delta = 2 \frac { h \beta}{H_0} \int_0^z \frac {(1+z) dz}{\sqrt{(\Omega_m +
\Omega_{dm})(1+z)^3 + \Omega_{de}}},
\eea
where $h$ is the energy density of the scalar field, $z$ is the
redshift factor, and $\Omega$'s are the cosmological 
density parameters. In terms of the density of the scalar field we can
write down the expression for the optical rotation as
\bea
\Delta \simeq \frac {M_p}{\omega \beta} \sqrt{ \rho} 
\times 5.6 \times 10^{43}~~~;~~~\rho=\frac {\omega^2 h^2 (1+z)^6}{2}
\eea
where $M_p$ is the Planck constant. In the above expression we consider
$z=0.4$ just because observational
data for radio galaxies and quasars have been analyzed in great detail for 
the redshift $z\geq 0.4$.
As one can see from the above expression, the optical rotation is 
crucially dependent upon the scale of alpha variation, coupling constant
$\beta$ and the energy density of the new scalar field.
In the next subsection we will investigate its impact on the
CMB polarization and constrain the value of the parameter 
$\beta$ in our model.

\subsection{Effect of birefringence on CMB anisotropy}
As we have already discussed, 
the CMB is one of the primary windows to peek into the 
early Universe. Recent CMB observations have reached
remarkable precision and proved to be consistent 
with the so-called standard model of cosmology.
With such high precision we can expect that
the CMB may provide additional information to constrain 
new physics beyond the standard model. 
A positive answer is expected from the study of CMB polarization. 
In the context of parity violating effects, there have already been
many studies \cite{ng}. These violations might also have a 
measurable imprint on the observed CMBP pattern, whose statistical 
properties are constrained by the assumption of symmetry conservation.

 It has been noted by several authors \cite{marc,lepora} 
that certain non-vanishing
multipole moment correlations between the temperature anisotropy and
polarization of the CMB could appear, if there exists parity violating
interaction in the photon sector. Such an interaction appears in 
our proposal in the framework of varying alpha theory. 
As is well-known, the angular distribution of the 
temperature anisotropy of the CMB can be expressed in terms of the  
expansion in spherical harmonics \cite{cowh}:
\begin{eqnarray}
{\frac {\Delta T} T}({\bf n})~=~\sum_{l,m}a^T_{lm}~Y^T_{lm}({\bf n})~.
\label{tani}
\end{eqnarray}
The polarization of the CMB is expressed in terms of a $2\times 2$
traceless symmetric tensor ${\cal P}_{ab}({\mathbf n})$ whose
components are the Stokes parameters. This tensor can be decomposed
into its irreducible `gradient' (or $E$) and `curl' (or $B$) parts
that have opposite spatial parities. The angular distribution of this
polarization tensor can thus be expressed in terms of the matrix
spherical harmonics as \cite{marc,lepora}

\begin{eqnarray}
{\cal P}_{ab}^E({\mathbf n})&~=~&\sum a_{lm}^E~Y_{lm,ab}^E({\mathbf 
n})~ \nonumber
\\
{\cal P}_{ab}^B({\mathbf n})&~=~&\sum a_{lm}^E~Y_{lm,ab}^B({\mathbf n})~.
\label{pol}
\end{eqnarray}
One defines the correlation of the multipole moment coefficients,
$a_{lm}^X~,~X = T,E,B$, as
\begin{eqnarray}
C_{l}^{XX'} \equiv \langle a_{lm}^X~a_{lm}^{X'} \rangle ~. 
\label{cor} 
\end{eqnarray}
Clearly, correlations such as
$C_l^{XX}$ as well as $C_l^{TE}$ all preserve P, while
correlations such as $C_l^{TB}$ and $C_l^{EB}$ are obviously P-violating, the appearance of which requires an explicitly P-violating interaction as mentioned earlier.
The optical activity described earlier implies that if a correlation
like $C_l^{TE}$ does indeed arise due to reionization or otherwise,
then the passage of the Thompson scattered photons through the scalar field $\phi$
background would produce the P-violating correlation term $C_l^{TB}$
through the rotation \cite{marc,lepora}:
\begin{eqnarray}
\label{eq:TC}
C_l'^{TB}&=& C_l^{TE}\sin 2 \Delta\\
\label{eq:GC}
C_l'^{EB}&=&\frac{1}{2}(C_l^{EE}- C_l^{BB})\sin 4 \Delta  
\end{eqnarray}
where the primed quantities are rotated and $\Delta$ is the 
rotation of the plane of polarization of light. We
clearly see that the effect of cosmic birefringence, which is
parity violating in nature, in our model can lead to
some nonvanishing correlations.

The recent high precession cosmological observations
put a tight constraints on the possible amount of optical 
rotation compare to the previous studies \cite{feng,maravin,kometsu}. 
The polarization
data from radio galaxies and quasars for the redshift between $z=0.425$ and 
$z=2.012$ gives the average value of $\Delta = -.6^0 \pm 1.5^0$.
On the other hand, the WMAP 7-years data \cite{kometsu} suggests that the
rotation angle of the polarization plane would be $\Delta = -1.1^0 \pm 1.3^0$.
That is, according to the WMAP polarization data there is no clear indication
for the parity-violating interaction in the photon sector. However, as we
have mentioned above, the most stringent constraint would come from the 
nonvanishing $TB$ and $EB$ correlations, whose values, as our model 
predicts, are different by a factor $\sin 2\Delta\sim 8\beta\delta\phi$. 
Since $\beta$ is
a free parameter to be fixed in our model, we need additional observational constraints to fix it. In the next section
we will discuss about the variation of $\alpha$ induced by our PCP
violating term. In principle this will help us 
fix the $\beta$. 
%In this paper we are not going to fixed those
%parameters comparing with the experimental results. We will report on this
%some where else.

\section{Effect of background electromagnetic field} \label{sec3}
Apart from the cosmological or astrophysical observations, there exist 
various laboratory-based experiments such as BFRT \cite{bfrt}, 
PVLAS \cite{pvlas},
Q\&A \cite{qa}, BMV \cite{bmv}, etc., which make use of the photon-to-scalar-field conversion
in the presence of a strong background magnetic or electric field for the indirect detection of new scalar fields.
In this regard different theoretical models based on the
dilaton-photon type coupling, $e^{-2 \phi} F_{ab}{F}^{ab}$,
or the standard QCD axion-photon type coupling, $\phi F_{ab}{\tilde F}^{ab}$, 
mediated by the background magnetic or electric field have
been considered extensively. In our present model we have 
employed both these terms in a single varying alpha
framework. As a first step, in this section we will try to do a qualitative 
analysis of our model under the background magnetic field. 
We want to emphasize here that this study
is important in the cosmological context as well. As we know,
at cosmological scales there exist
background magnetic fields. These cosmic magnetic fields may have a 
significant effect on the CMB polarization in addition to the scalar coupling effect that we describe in this paper.
The polarization of CMB is known to have encoded the information of 
early Universe specifically that of the inflationary epoch. The possibility of additional CMB polarizations induced by some other external field would undoubtedly complicates the issue and it must be clarified. 
With this motivation in mind, we calculate the effect 
of background electromagnetic field on the rotation plane of
polarization. In terms of the vector potential, the main equations of our 
interests are
\bea
&&(\nabla^2  + \varpi^2) {\bf A}_x = 4 i \beta {\bf B}_0 \varpi \phi, \\
&&(\nabla^2 + \varpi^2) {\bf A}_y =- 2{\bf B}_0 \pr_z \phi, \\
&&(\nabla^2 + \varpi^2) {\bf A}_z = 2 {\bf B}_0 \pr_y \phi, \\
&&(\nabla^2 + \varpi^2) \phi =\frac {2 {\bf B}^2_0}{\omega} \phi 
- \frac {2 {\bf B}_0}{\omega}
(\pr_y A_z - \pr_z A_y) - 
\frac {4 i \beta {\bf B}_0 \varpi}{\omega} A_x, 
\eea
in the presence of background magnetic field ${\bf B}_0$ in the x-direction.
Because of the smallness of the effect, we consider only the linear order equations
for the scalar-photon system. 
In the above derivation we used the gauge condition, $\nabla \cdot {\bf A} = 0$,
and specified the scalar potential: ${\bf A}_0 = 0$. $\varpi$ is the frequency of the fields.  
Let the propagation direction of the electromagnetic wave be 
orthogonal to the external magnetic field ${\bf B}_0$, say in $z$-direction. We
then write
\bea
{\bf A}(z,t) = {\bf A}^0 e^{- i \varpi t + i k z}~~~ ;~~~
 \phi(z,t)=\phi^0 e^{- i \varpi t + i k z} . 
\eea
As is clear from the above ansatz, the equation for $A_z$ is no longer coupled
with $\phi$. From the other three equations for $A_x,A_y,\phi$,
consistency condition leads to three roots for the frequency $\varpi$ as 
follows
\bea
\varpi^2  &=& k^2~~~,~~~\varpi_{\pm}^2 = k^2 + \delta_{\pm}\\
\delta_{\pm}&=& \frac {{\bf B}_0^2 } {\omega}(1 + 8 \beta^2)  \pm
\sqrt{\frac {{\bf B}_0^4 } {\omega^2}(1 + 8 \beta^2)^2 +
\frac {4{\bf B}_0^2 k^2}
{\omega} (1+4\beta^2)}.
\eea  
To establish the connection with the experimental set up, 
we can consider the initial ($t = 0,x=0$) electromagnetic field to be 
linearly polarized and making an angle with the external magnetic field
${\bf B}_0$ , so that 
\bea
{\bf A}_x(z=0,t=0) = \alpha_1= \cos \alpha~~;~~  
{\bf A}_y (z=0,t=0)= \alpha_2= \sin\alpha ~~;~~ \phi(z=0,t=0)= 0.
\eea
With these boundary conditions, we can have a unique solution like
\bea
&&{\bf A}_x = (a_x e^{-i \varpi t} + b_x e^{-i \varpi_{+} t}+ 
c_x e^{-i \varpi_{-} t})e^{i k z} \nno\\
&&{\bf A}_y=( a_y e^{-i \varpi t} + b_y e^{-i \varpi_{+} t}+ 
c_y e^{-i \varpi_{-} t})e^{i k z}  \nno\\
&&\phi = \phi_0 ( e^{-i \varpi_{+} t} -  e^{-i \varpi_{-} t}) e^{i k z}  ,
\eea
where
\bea
b_x &=&- \frac {2 \beta \varpi_{+}}{k} b_y = \frac {2 \beta \varpi_{+}}{k}
\frac {(-k^2+\varpi_{-}^2)}{(-k^2+\varpi_{+}^2)} c_y = -\frac {\varpi_{+}}
{\varpi_{-}} \frac {(-k^2+\varpi_{-}^2)}{(-k^2+\varpi_{+}^2)} c_x
= \frac {4 i \beta {\bf B}_0 \varpi_{+}}{(-k^2+\varpi_{+}^2)} \phi_0
\nno\\
a_y &=& \frac {2 \beta \varpi}{k} a_x = \alpha_2 + \frac{k} {2 \beta
\varpi_- }
\left[\frac {\varpi_{+}^2-\varpi_{-}^2}{-k^2+\varpi_{+}^2}
\right] c_x \nno \\
c_x &=& \frac {1}{{\cal F}}\left(\alpha_1 - \frac {k \alpha_2}
{2 \beta \varpi}\right) \nno \\
{\cal F}&=& \frac {4 \beta^2 (\varpi \varpi_{-}
(-k^2 + \varpi_{+}^2)-\varpi \varpi_{+}
(-k^2 + \varpi_{-}^2))+ k^2(\varpi_{+}^2-
\varpi_{-}^2)}{4 \beta^2 \varpi \varpi_{-}(-k^2+\varpi_{+}^2)}
\eea

While traversing through the region of external magnetic field, after $t=L$
the resulting interaction causes the wave solution to have a modified 
amplitude of the form 
\bea \label{sol_l}
&&{\bf A}_x = a_x e^{-i \varpi L} + b_x e^{-i \varpi_{+} L}+
c_x e^{-i \varpi_{-} L}, \\
&&{\bf A}_y= a_y e^{-i \varpi L} + b_y e^{-i \varpi_{+} L}+
c_y e^{-i \varpi_{-} L}, 
\eea
From the above set of expressions, we see that the vector potential
describes an ellipse with the major axis at an angle
\bea \label{rot}
\theta \simeq \tan^{-1}\left(\frac {\alpha_2}{\alpha_1}\right) +
\frac {\sin (2 \alpha)} {4}\left( \frac {{\cal L}}
{\cos^2(\alpha)} - \frac {{\Gamma}} {\sin^2(\alpha)} \right)
\eea
where
\bea
&&{\cal L} = 2 a_x b_x \sin^2(\frac {\Delta_{+}}{2}) + 2 a_x c_x
\sin^2(\frac {\Delta_{-}}{2})+2 c_x b_x \sin^2(\frac {\Delta}{2}), \nno\\
&&{\Gamma} = 2 a_y b_y \sin^2(\frac {\Delta_{+}}{2}) + 2 a_y c_y
\sin^2(\frac {\Delta_{-}}{2})+2 c_y b_y \sin^2(\frac {\Delta}{2}), \nno\\
&&\Delta_{+} =( \varpi_{+} -\varpi)L~~~;~~~\Delta_{-} =
(\varpi_{-} -\varpi)L~~~;~~~
\Delta = (\varpi_{+} -\varpi_{-})L
\eea
Now, eq.\ref{rot} yields the expression for the
optical rotation of the plane of polarization as
\bea
\delta = \frac {\sin (2 \alpha)} {4}\left( \frac {{\cal L}}
{\cos^2(\alpha)} - \frac {{\Gamma}} {\sin^2(\alpha)} \right)
\eea

This is the quantity that establishes the direct connection with the 
experimental data. The similar analysis can also be made for the background
electric field as well. In our forthcoming paper we will consider a more
detailed analysis of the background electromagnetic field effect 
on the scalar-photon mixing and its effect on the various
laboratory as well as cosmological experiments. 
So far we have studied the effect of the scalar field on the polarization of
the electromagnetic wave under various conditions that may arise in a 
laboratory or cosmological settings. In the next section we will consider the 
change of the scalar field or fine structure constant under the 
background cosmological evolutions.

\section{Varying $\alpha$ cosmology} \label{sec4} 
The effect of cosmic evolution on the variation of the fine structure
constant in the framework of the variation of a scalar field $\phi(x)$
 has been extensively studied \cite{bsm,bsm1,barrow}. This has 
been referred to as the Bekenstein-Sandvik-Barrow-Magueijo (BSBM) theory.
Here we only analyze the variation of $\alpha$ induced by the PCP
 violating effect. As we have already mentioned,
the effective time varying fine structure constant is
\begin{eqnarray}
\alpha(t) ~=~ e^{2\phi(t)}~ \label{alph}.
\end{eqnarray}
In the subsequent analysis we will switch over to the usual cosmic time. 
The fractional variation of $\alpha$ then becomes
\bea
\frac {\Delta \alpha} {\alpha(t_0)} ~=~ \frac {\alpha(t_0) -\alpha(t)}
{\alpha(t_0)} = 1 - e^{2[\phi(t)-\phi(t_0)]}
~\approx ~ 2 ~[\phi(t_0) - \phi(t)] = 2 \Delta 
\phi(t), 
\eea
where $t_0$ refers to the present epoch. The observational upper limit of the time 
variation of the fine structure constant \cite{webb} then puts a constraint on the variation of the scalar field, 
\be
\frac {|\Delta \alpha|} {\alpha(t_0)}  \simeq 10^{-5} .
\ee
In order to further constrain our model parameters we need to know 
the nature of solution for the scalar field $\phi(t)$. We will do so in the subsequent subsections.

\subsection{General analysis}
In this section we study the cosmological evolution of the scalar field
during the various phases of the Universe evolution history.
In the cosmological setting the equation of motion is 
\begin{equation}
G_{\mu \nu }= \frac 1 {M_p^2} \left(\langle T_{\mu \nu }^{mat}\rangle
 ~+~ T_{\mu \nu }^{\Phi_H} ~+~ 
e^{-2 \phi} \langle T_{\mu \nu}^{em} \rangle \right) .
\end{equation}
The average $\langle \cdots \rangle$ denotes a statistical 
average over the current state of the Universe. The electromagnetic field equation becomes
\be \label{electro}
\nabla_{\mu}[e^{-2 \phi}(\langle F^{\mu\nu}\rangle 
+ \beta \langle \tilde{F}^{\mu\nu})\rangle ] = 0,
\ee 
while variation with respect to the $\phi $ field gives the cosmological evolution 
for the field: 
\begin{equation}
\Box \phi  = \frac {e^{-2\phi}}{2 \omega}
\left[-~\langle F_{\mu \nu }F^{\mu \nu}\rangle ~+~ 
\beta~\langle F_{\mu \nu }{{\tilde F}^{\mu \nu}}\rangle\right].  
\label{boxPhi}
\end{equation}
 For our future convenience we use the notation ${\cal L}_{em} =- \frac 1 4 
F_{\mu \nu }F^{\mu \nu}$. 

In the standard electrodynamics both terms on the right-hand side
of Eq.(\ref{boxPhi}) vanish. 
%that the term $\langle F_{\mu \nu }F^{\mu \nu}\rangle \simeq 
%\langle({\bf E}^2-{\bf B}^2)\rangle$ on the right hand side 
%vanishes for pure radiation. On the other hand, clearly this is not for the second term  
%$\langle F_{\mu \nu }{{\tilde F}^{\mu \nu}}\rangle\ \simeq 
%\langle {\bf E}\cdot {\bf B}\rangle$.
The PCP-violating time variation of $\phi$, and therefore that of $\alpha$, causes the cosmic birefringence which in turn breaks the orthogonality properties of electromagnetic field, and as a result the term $\langle F_{\mu \nu }{{\tilde F}^{\mu \nu}}\rangle\ \simeq 
 \langle {\bf E}\cdot {\bf B}\rangle$ can in principle be nonvanishing during the radiation epoch. 
We emphasize that this particular effect on the $\alpha$ variation was not present in the original BSBM theory. The other known contribution to the variation of $\alpha$ 
comes from nearly pure electrostatic or magneto-static energy of the matter field. 
As has been extensively discussed in Refs.\cite{bek2,bsm,bsm1},
the nonrelativistic matter contributes to the right-hand side of Eq.~(\ref{boxPhi}) 
through the spatial variation of the Coulombic mass. This contribution
is parametrized by the ratio $\zeta_m={\cal L}_{em}/\rho$, where
$\rho$ is the energy density and ${\cal L}_{em}\approx E^2/2$ 
for baryonic matter. BBN infers an approximate value for the baryon
density of $\Omega _B\approx 0.03$ with a Hubble parameter $h_0\approx 0.6
$, implying $\Omega _{CDM}\approx 0.3$. So, $\zeta_m$ depends strongly 
on the nature of the dark matter and can be either positive or negative 
with a modulus between $0$ and $ 1$.

Assuming a spatially-flat, homogeneous and isotropic Friedmann metric with
expansion scale factor $a(t)$, 
\be
ds^2 = -dt^2 + a(t)^2 (dx^2 +dy^2 + dz^2),
\ee
we obtain the Friedmann equation 
\begin{eqnarray}
\left( \frac{\dot{a}}a\right)^2 = \frac{1 }{3 M_p^2}
\left[ \rho_m\left\{ 1+ e^{- 2 \phi}\zeta_m \right\} + 
e^{-2 \phi} \rho_r + \rho _{\phi}\right] + \frac {\Lambda} 3
\label{fried1}
\end{eqnarray}
where $\Lambda $ is a constant cosmological vacuum energy density and 
$\rho_{\phi} =\frac 1 2 [\dot{\phi}^2+ V(\phi)]$. For the scalar field we get
\begin{equation}
\ddot{\phi} + 3H\dot{\phi} =\frac {e^{-2\phi}} 
{\omega} [ -2 \zeta _m \rho _m  + 
\frac 4 {a^3}  \langle{\bf E}\cdot{\bf B} \rangle], 
\label{psiddot}
\end{equation}
where $H\equiv \dot{a}/a.$ The conservation equations for the
noninteracting radiation and matter densities $\rho _r$ and $\rho _m$,
respectively, are
\begin{eqnarray}
\dot{\tilde{\rho}}_m+3H \tilde{\rho}_m &=&0, \\
\pr_t ( e^{-2 \phi} \rho_r) + 4H e^{-2 \phi} \rho_r &=& 0 , 
\label{dotrho}
\end{eqnarray}
where $\rho_r$ is the radiation energy density. From the last equation one 
finds $\tilde{\rho}_r \equiv e^{-2  \phi} \rho _r \propto a^{-4}$, while
the solution for the matter density is $\tilde{\rho}_m =\{1+ e^{ -2
\phi}\} \rho_m \propto a^{-3}$.
Equations (\ref{fried1}-\ref{dotrho}) govern the Friedmann universe with a
time-varying fine-structure constant $\alpha(t)$. 
They depend on the choice of the parameters $\zeta_m/{\omega}$ and 
$\beta/{\omega}^2$.
In general it is difficult to solve the Eqs.(\ref{fried1},\ref{psiddot}).
Since the effect of the new scalar field is expected to be very small on the
background cosmological evaluation, we will try to solve the scalar field evolution equation in the leading order approximation with 
the standard Hubble expansion included.

\subsection{Evolution of scalar field in different cosmological era}
In this section we analyze the evolution of the scalar field in the various cosmological
eras. For simplicity as well for analytical purposes, 
we will ignore the potential term of the field.

\subsubsection{The Radiation dominated era}
We here show that during the radiation era there exists
a contribution to the variation of $\alpha$ 
through PCP violating term as opposed to the usual Bekenstein
theory. In this era The Friedmann equation is
\begin{eqnarray}
\left(\frac{\dot{a}}a\right)^2 = \frac{1 }{3 M_p^2}\left[ 
e^{- 2 \phi} \rho_r   +  \frac 1 2 \dot{{\phi}}^2 \right],
\label{radiation}
\end{eqnarray}
while the equation for the scalar field becomes
\begin{equation}
\frac d {d t} (\dot{\phi} a^3 ) = e^{-2\phi} \frac {4 \beta } 
{\omega} \langle{\bf E}\cdot{\bf B} \rangle.
\end{equation}
As we have discussed before, the average value
of a radiation kinetic Lagrangian in pure radiation does not contribute to the $\alpha$ evolution.
In order to solve the above equation for the scalar field, we need to
know the average value of the PCP violating term in the action.
However, we observe from Eq.(\ref{electro})
that in the plane wave limit the essential equation for our study is 
\be \label{EB}
\partial_0(a {\bf E} \cdot {\bf B}) = a {\bf E}\cdot(\nabla\cdot{\bf E})
+ \frac 1 a {\bf B}\cdot(\nabla\cdot{\bf B})+
 {\dot{{\phi}}}(2 a {\bf E} \cdot {\bf B} - 4 \beta {\bf B} \cdot {\bf B}).
\ee
It is clear from the above equation that ${\bf E}$ and
${\bf B}$ are not perpendicular to each other due to varying
fine structure constant. In the plane wave limit, we can ignore the first two terms because
$\kappa\cdot {\bf B}=\kappa\cdot {\bf E}=0$, where $\kappa$ is the wave
propagation direction.
We then find
\be
a \langle{\bf E}\cdot{\bf B} \rangle={\langle {\bf B}\cdot {\bf B}
 \rangle}
\left(2{ \beta} +  \theta e^{2\phi}\right) ,
\ee
where $\theta$ is the integration constant. Equation (\ref{EB}) is a first
order differential equation in time. Therefore we can chose the
initial condition to be orthogonal i.e.
${\bf E}\cdot{\bf B}=0$ such that $\theta = -2 \beta e^{-2\phi_0}$,
where the initial value of $\phi$ is taken to be $\phi(t_i)= \phi_0$.
The parameter $\beta$
of our model therefore plays the main role in
breaking the orthogonality of the electromagnetic field.
The evolution equation for $\phi$ now becomes
\be \label{phiEB}
\frac d {d t} (\dot{\phi} a^3 ) =\frac {8\beta^2
\langle{\bf B} \cdot{\bf B} \rangle}
{a \omega}\left(e^{-2\phi} - e^{-2\phi_0}\right).
\ee
We see that the variation of $\alpha$ depends quadratically in $\beta$.
As we have mentioned before, in order to solve the above set of 
equations analytically, we invoke a self-consistent
approximation which has been employed in \cite{bsm}. 
The basic strategy of this approximation is that it invokes the
background solution for the cosmological scale factor
in the equation that governs the scalar field evolution. 
This is justified since at a late stage
in the radiation era, the energy of the scalar field should fall faster than
that of the radiation. 

Specifically, we assume that the scale factor is 
$a(t)=t^{1/2}$ for the radiation era. Changing the variable to
$x=\frac 1 2 \ln(t)$, we find that Eq.(\ref{phiEB}) becomes
\begin{equation} 
\phi^{\prime \prime }+\phi ^{\prime }= {\cal A} 
\left(e^{-2\phi} - e^{-2\phi_0}\right) , 
\end{equation}
where $^{\prime }\equiv d/dx$ and
\[{\cal A} = \frac {8\beta^2
\langle{\bf B} \cdot{\bf B} \rangle} {\omega}\geq 0. 
\]
The above equation is very difficult to solve analytically. In order to
get an analytic expression, let us assume that field variation
is very small. Under this approximation we can write down
\bea \label{phiEB1}
\phi^{\prime \prime }+\phi ^{\prime }+2 {\cal A}(\phi -\phi_0) =0 ,
\eea
Equations (\ref{phiEB1}) can be solved exactly for the
varying fine structure constant:
\bea
\phi =\phi_0 + {\cal C}_1 x^{-{\alpha_+}} + {\cal C}_2 x^{-{\alpha_- }}
 ~~;~~\alpha_{\pm}
= \frac 1 4 \left(1 \pm \sqrt{1- 8 {\cal A}}\right).
\eea 
In the above
discussion for the orthogonality, we chose the initial value 
of the scalar field to be $\phi_0$, which fixes ${\cal C}_2 =- {\cal C}_1 
x_i^{\delta \alpha} $, where $\delta \alpha = \alpha_- - \alpha_+$. 
Other constant can be fixed by matching the value of 
a fine structure constant at the matter-radiation equality epoch.
With the above solution, the expression for the 
fine structure constant during the radiation dominated era is
\begin{equation} 
\alpha \sim  \exp \Big[\phi_0+ 2 {\cal C}_1 t^{{-\alpha_+}} + 
2 {\cal C}_2 t^{-{\alpha_- }}\Big]. \label{ald2}
\end{equation}
As we have already mentioned before, in the above solution
the back reaction of the scalar field has not been considered in the 
background evolution. The standard radiation dominated
cosmic expansion is therefore unperturbed.  
To check the validity of this approximation, 
we compare the leading order behavior of the energy densities of the 
radiation and the scalar field:
\begin{equation}
e^{- 2 \phi} \rho _r \propto a^4 = \frac 1 {t^2} ,~~~~ 
\rho_{\phi} = \frac {\omega} 2 {\dot {\phi}}^2 
\propto \frac {{{\cal C}_1}^2} {t^2}\frac 1 {\ln(t)^{2\alpha_++2}}~,~
 \frac {{{\cal C}_2}^2} {t^2}\frac 1 {\ln(t)^{2\alpha_-+2}}
\end{equation}
As is clear from the above two expressions for the energy densities,
the $\dot{\phi}^2$ term falls off faster than the radiation energy
density as $t\rightarrow \infty $.
From Eq.(\ref{ald2}) we see that depending on the boundary conditons
$\alpha $ can decrease or increase with time in the 
radiation dominated epoch. 
The change of $\alpha
$, on the other hand, is controlled by the average energy density 
of the radiation, ${\cal A}$, as well as the PCP violating 
coupling, $\beta$. 

In the context of the subsequent cosmic expansion, the new PCP 
violating term in our Lagrangian does not contribute to the evolution 
of the scalar field ${\phi}$.
Therefore the corresponding variation of the alpha has the same evolution in 
the subsequent matter and dark energy dominated eras. 
This has been extensively discussed in Refs. \cite{bsm,bsm1,barrow}.
 
\section{Conclusions} \label{con}
We have constructed a parity and charge-parity (PCP)
violation model within the framework
of the varying alpha theory, popularly known as BSBM theory \cite{bek2,bsm1}.
The origin of this violation in our model is the time variation of the charge, which is the basic assumption of this framework.
One of the main motivations for this model is to search for new physics constrained by the present-day high precession data from cosmological observations.  
After constructing our model, we have calculated various relevant effects such as the cosmological birefringence, which has already become a standard observational parameter in CMB as
well as in radio galaxy and quasar spectra observations.
Although until now there has been no positive observational evidence of this 
parity-violating 
effect, future experiments with ever improved precession
may hopefully help us identify this notion beyond the standard model.
Our model also predicts that this new contribution to the 
fine-structure constant
variation is effective mainly in the radiation dominant era. In other eras,
the variation is essentially the same as those extensively discussed 
in the literature 
\cite{bsm,bsm1,barrow}. 
Because of that, BBN (Big Bang Nucleosynthesis) becomes the main 
observational window to constrain the evolution of the
PCP violating varying fine structure 
constant. The electromagnetic coupling constant 
plays a very significant role in the nuclear abundance 
of our Universe. It happens that in our model the fine structure constant has
a power law time variation during the radiation dominant era, whereas 
in the standard BSBM model it remains almost constant. So
BBN should give us a strong constraint on the parity violating parameter.

As is well known, BBN needs three 
essential input parameters which are the 
neutron-proton mass difference, $\Delta m$, the neutrino life time, $\tau_n$,
and the nuclear reaction rates. All of these parameters are directly or
indirectly depending upon the fine structure constant. There have been
extensive studies on constraining the fine structure constants through
the light element abundance. The most updated bound on the total variation of  
the alpha is $-0.007 \leq \delta\alpha/\alpha_0
\leq 0.017$ at $95\%$ C.L. \cite{Dent}. In order to constrain the
parity-violating parameter $\beta$, we need to know 
the amount of variation of the fine structure constant 
after the radiation dominated epoch. To accomplish this, 
we need one more constraint deduced from a later time in cosmic evolution.

CMB anisotropy is another powerful 
tool to constrain the possible 
variation of a fine-structure constant from the matter dominated epoch to the
present epoch.  
Variation of the alpha during the matter dominated epoch before CMB 
would change the time of recombination and the acoustic horizon
associated with the photon-electron decoupling.The most updated
bound on the variation of the fine-structure constant has been reported in
\cite{landau} by using the latest WMAP 7-year data, and that is  
$-0.005 \leq \delta \alpha/\alpha_0\leq 0.008$ at $95\%$ C.L.
By comparing the aforementioned two different bounds 
on the alpha variation deduced from two different cosmological time scales, 
it may be possible to constraint the PCP violating parameter 
$\beta$ of our model. As a rough estimate, we take the 
difference between these two constraints and find the bound for the radiation-dominant era:
$-0.002 \leq \delta \alpha_{rad}/\alpha_0
\leq 0.009$, where ${\delta \alpha}_{rad} 
\approx \left(2 {\cal C}_1 t_{eq}^{{-\alpha_+}} + 
2 {\cal C}_2 t_{eq}^{-{\alpha_- }}\right)$; $t_{eq}$ is 
time of radiation-matter equality during the cosmic evolution.

Apart from the constraints deduced from cosmological and astrophysical observations, 
we have also done some qualitative analysis on the amount of the optical 
rotation due to background electromagnetic fields. Because of 
the existence of cosmic-scale magnetic fields in our Universe,
polarization of the CMB photons may be sizable due
to their coupling to the scalar field. With these considerations 
in mind, we believe that there exists experimental
windows through which the validity of our model
or the constraints of its parameters can be verified.
As a first step, we have focused on establishing the qualitative behavior of 
our model in the present paper, but we did not investigate the 
observational constraints on its parameters. 
We hope to study this in more details in the future.  
 
\vspace{.1cm}

{\bf Acknowledgement}\\
This research is supported by Taiwan National Science Council under Project No. NSC
97-2112-M-002-026-MY3, by Taiwan's National Center
for Theoretical Sciences (NCTS), and by US Department
of Energy under Contract No. DE-AC03-76SF00515.


\begin{thebibliography}{99}

\bibitem{carroll} S. M. Carroll, 
Phys. Rev. Lett. {\bf 81}, 3067 (1998) [arXiv:astro-ph/9806099]. 

\bibitem{marc} A. Lue, L. M. Wang and M. Kamionkowski, 
Phys. Rev. Lett. {\bf 83}, 1506 (1999) [arXiv:astro-ph/9812088].

\bibitem{contaldi} C. R. Contaldi, J. Magueijo and L. Smolin, 
Phys. Rev. Lett. {\bf 101}, 141101 (2008) [arXiv:0806.3082 [astro-ph]].

\bibitem{soda} T. Takahashi and J. Soda, 
Phys. Rev. Lett. {\bf 102}, 231301 (2009) [arXiv:0904.0554 [hep-th]].

\bibitem{debu} D. Maity, P. Majumdar, S. SenGupta, JCAP, {\bf 0406}, 
005 (2004); P. Majumdar, Mod. Phys. Lett. {\bf A19}, 1319 (2004) 
[hep-th/0105122]; K.R.S. Balaji, R.H.
 Brandenberger, D.A. Easson, JCAP, 0312, 008, (2003).

\bibitem{gammow} P. Jordan, Naturwiss. {\bf 25}, 513 (1937); 
Z. f. Physik {\bf 113}, 660 (1939); E. Teller, Phys. Rev. {\bf 73}, 
801 (1948); 
K. P. Stanyukovich, Dokladii Akad. Nauk SSSR {\bf 147}, 1348 (1962) 
[Sov. Phys. Doklady 7, 1150 (1963)]; G. Gamow, Phys. Rev. Lett. {\bf 19}, 
759 (1967); R. H. Dicke, 
The Theoretical Significance of Experimental Relativity 
(Gordon and Breach, New York 1965); R. H. Dicke, Science {\bf 129}, 621 (1959).

\bibitem{bek2}  J.D. Bekenstein, Phys. Rev. D 25, 1527 (1982).

\bibitem{bsm} H. B. Sandvik, J. D. Barrow and
J. Magueijo, Phys. Rev. Lett. {\bf 88}, 031302
(2002), Phys. Rev. D {\bf 65}
, 123501 (2002), Phys. Rev. D {\bf 66}, 043515 (2002), 
Phys. Lett. B {\bf 541}, 201 (2002), J.D. Barrow
and D. Mota, Class. Quant. Grav. {\bf 19}, 6197 (2002).

\bibitem{bsm1} J. D. Barrow, H. B. Sandvik and
J. Magueijo, Phys. Rev. D {\bf 65}, 063504 (2002).
 
%\bibitem{albrecht} A. Albrecht and J. Magueijo, Phy. Rev, D {\bf 59}, 043516 (1999).
\bibitem{barrow} K. A Olive and Maxim Pospelov, hep-ph/0110377;
T. Chiba and K. Kohri, Prog. Theor. Phys. {\bf 107}, 631 (2002);
J. P. Uzan, Rev. Mod. Phys. {\bf 75}, 403 (2003); D. S. Lee, W. Lee and K. W. Ng, Int. J. Mod. Phys. D {\bf 14}, 335 (2005).

\bibitem{murphy}  M.T.Murphy et. al, MNRAS, 327, 1208 (2001) .

\bibitem{webb}  J.K. Webb, V.V. Flambaum, C.W. Churchill, M.J. Drinkwater \&
J.D. Barrow, Phys. Rev. Lett. 82, 884 (1999).
\bibitem{webb2}  J.K. Webb et al, Phys. Rev. Lett. 87, 091301 (2001).


\bibitem{maroto} A. Dobado and A. Maroto, Mod. Phys. Lett. A {\bf 12}, 3003
(1997).

\bibitem{mas} D. Maity and S. SenGupta, Class. Quant. Grav. 
{\bf 21}, 3379 (2004) [arXiv:hep-th/0311142]. 

\bibitem{dmssssg} D. Maity, S. SenGupta and S. Sur, Phys. Rev. D
{\bf 72}, 066012 (2005) [hep-th/0507210]; S. Kar, P. Majumdar,
S. SenGupta and S. Sur, Class. Quant. Grav. {\bf 19}, 677 (2002)
[hep-th/0109135].

\bibitem{ng} G. C. Liu, S. Lee and K. W. Ng, Phys. Rev. Lett. {\bf 97}, 161303 (2006); M. Li {\it etal}, Phys. Lett. B {\bf 651}, 357 (2007); M. Shimon {\it etal}, Phys. Rev. D {\bf 77}, 083003 (2008); J. Q. Xia {\it etal}, Astron. Astrophys. {\bf 483}, 715 (2008); F. Finelli and M. Galaverni, Phys. Rev. D {\bf 79}, 063002 (2009); J. Q. Xia {\it etal}, Astrophys. J. {\bf 679}, L61 (2008); M. Pospelov, A. Ritz and C. Skordis, Phys. Rev. Lett. {\bf 103}, 051302 (2009); M. Li  and X. Zhang, Phys. Rev. D {\bf 78}, 103516 (2008); J. Q. Xia {\it etal}, Phys. Lett. B {\bf 687}, 129 (2010); L.  Pagano, {\it etal}, Phys. Rev. D {\bf 80}, 043522 (2009).


\bibitem{lepora} N. Lepora, gr-qc/9812077.

\bibitem{cowh} M. White and J. Cohn 2002, TACMB-1, arXiv:astro-ph/0203120.

\bibitem{feng} B. Feng {\it et al}., Phys. Rev. Lett. {\bf 96}, 
221302 (2006) [arXiv:astro-ph/0601095].

\bibitem{maravin} P. Cabella, P. Natoli and Joseph Silk,
Phys. Rev. D {\bf 76}, 123014 (2007);  T. Kahniashvili, R. Durrer and Y. Maravin,
Phys. Rev. D {\bf 78}, 123009 (2008) [arXiv:0807.2593 [astro-ph]]; 
J. Q. Xia {\it et al.}, Astrophys. J. {\bf 679}, L61 (2008) 
[arXiv:0803.2350 [astro-ph]]; J. Q. Xia {\it et al.}, Astron. Astrophys. 
{\bf 483}, 715 (2008) [arXiv:0710.3325 [hep-ph]]; E. Y. Wu {\it et al.} 
[QUaD Collaboration], Phys. Rev. Lett. {\bf 102}, 161302 (2009) 
[arXiv:0811.0618 [astro-ph]]; L. Pagano {\bf et al.}, Phys. Rev. D {\bf 80},
043522 (2009) [arXiv:0905.1651 [astro-ph.CO]].

\bibitem{kometsu} E. Komatsu {\it et   al.}, [WMAP Collaboration], 
arXiv:1001.4538 [astro-ph.CO] 


\bibitem{bfrt} Y. Semertzidis {\it et al.} [BFRT Collaboration],
Phys. Rev. Lett. {\bf 64}, 2988 (1990), R. Cameron 
{\it et al.} [BFRT Collaboration], Phys. Rev. D. {\bf 47}, 3707 (1993).

\bibitem{pvlas} E. Zavattini {\it et al.} [PVLAS Collaboration], Phys. Rev. Lett. {\bf 96} (2006) 110406,
[arXiv:hep-ex/0507107].

\bibitem{qa} S. J. Chen, H. H. Mei and W. T. Ni [Q \& A Collaboration], arXiv:hep-ex/0611050.

\bibitem{bmv}C. Rizzo for the [BMV Collaboration], 2nd ILIAS-CERN-CAST Axion Aca:demic Training 2006,
http://cast.mppmu.mpg.de/
\bibitem{Dent} T. Dent, S. Stern and C. Wetterich, Phys. Rev. {\bf D76},
063513 (2007).
\bibitem{landau} S. J. Landau and C. G. Scoccola, arXiv:1002.1603 [astro-ph.CO] 

\end{thebibliography}
\end{document}